# Pressure-induced hypercoordination of iodine and dimerization of I$_2$O$_6$H in strontium di-iodate hydrogen-iodate (Sr(IO$_3$)$_2$HIO$_3$)

D. Errandonea,[1,*] H. Osman,[1,2] R. Turnbull,[1] D. Diaz-Anichtchenko,[1] A. Liang,[3] J. Sanchez-Martin,[1] C. Popescu,[4] D. Jiang,[5] H. Song,[5] Y. Wang,[5] F.J. Manjon[2,*]

[1]Departamento de Física Aplicada-ICMUV, MALTA-Consolider Team, Universidad de Valencia, Dr. Moliner 50, Burjassot, Valencia 46100, Spain

[2]Instituto de Diseño para la Fabricación y Producción Automatizada, MALTA Consolider Team, Universitat Politècnica de València, 46022 València, Spain

[3]Centre for Science at Extreme Conditions and School of Physics and Astronomy, University of Edinburgh, Edinburgh EH9 3FD, United Kingdom

[4]CELLS-ALBA Synchrotron Light Facility, Cerdanyola, Barcelona 08290, Spain

[5]School of Materials Science and Engineering, Peking University, Beijing 100871, China

*Corresponding authors: daniel.errandonea@uv.es and fjmanjon@fis.upv.es

**Abstract:** In this work, we report evidence of pressure-induced changes in the crystal structure of Sr(IO$_3$)$_2$HIO$_3$ connected to changes the coordination of the iodine atom and the of the configuration of HIO$_3$ and IO$_3$ units. The changes favor iodine hypercoordination and happen in two steps on sample compression. Firstly, at 2.5 GPa, [HIO$_3$]·[IO$_3$] complexes are formed, and secondly, at 4.5 GPa, these complexes form dimers of [HIO$_3$]·[IO$_3$]·[IO$_3$]·[HIO$_3$]. The evidence is obtained from a combined experimental and theoretical study performed up to 20 GPa. Synchrotron powder X-ray diffraction, Raman spectroscopy, and optical-absorption experiments have been complemented with density-functional theory calculations, including the study of the topology of the electron density. The changes observed in the crystal structure are related to the transformation of secondary (halogen) bonds into electron-deficient multicenter bonds. The paper also discusses the effect of pressure on the compressibility of the Sr(IO$_3$)$_2$HIO$_3$ crystal structure, its phonons, the electronic band gap, and the refractive index. Sr(IO$_3$)$_2$HIO$_3$ was found to be highly compressible with an anisotropic compressibility. The softening of the internal I-O vibrations of IO$_3$ units was also observed, together with a decrease of the band-gap energy (from 4.1 eV at 0 GPa to 3.7 eV at 20 GPa), a band-gap crossing, and a change in the topology of the band structure, with Sr(IO$_3$)$_2$HIO$_3$ transforming from a direct gap semiconductor at 0 GPa to an indirect gap semiconductor beyond 6 GPa.

## 1. Introduction

Pentavalent iodine (I[5+]) has a stereochemically active lone-electron pair (LEP). The presence of these non-bonding valence electrons leads to the existence of a diversity of crystal structures in iodates [1]. Multiple iodate compounds have been reported, with more than 450 entries registered for iodates with pentavalent iodine, in the Inorganic Crystal Structure Database (ICSD). The presence of trigonal [IO$_3$]$^-$ pyramids formed by the tricoordinated iodine atoms, is a shared characteristic of the crystal structure of most iodates. Another characteristic feature of iodates is the existence of secondary interactions of I atoms with the second-neighbor O atoms. These interactions are cataloged in the literature as halogen bonds, according to current IUPAC nomenclature, in which the LEPs are involved [2]. The LEP of iodine provides iodates and iodides with multiple interesting characteristics that favor their application in different technologies. They include birefringence, ferroelectricity, piezoelectricity, second-harmonic generation, and barocaloric properties [3-9]. Because of the above-described facts, iodates materials are increasingly receiving attention from materials scientists, solid-state physicists, and chemists [1].

Over the last five years, large attention has been devoted to high-pressure (HP) studies of metal iodates [1, 10]. These studies were motivated not only to contribute to the development of applications of iodates but also to fundamental research. Due to the presence of the iodine LEP in the [IO$_3$]$^-$ unit, iodates have been discovered to be highly compressible [11]. It has been found also that compression of metal iodates can trigger other phenomena, such as: symmetry-preserving phase transitions [11, 12]; the softening of phonons associated with the internal vibrations of the iodate molecule [12, 13]; and, a non-linear behavior of the pressure dependence of the band-gap energy, which is governed by the pressure-induced changes in the I-O bond lengths [14]. There are additional interesting properties revealed by high-pressure research, including include the discovery that iodates can exhibit intrinsic zero-linear and zero-area compressibilities [15] or even a negative linear compressibility in at least one of the crystallographic axes [16]. Another interesting observation of recent HP studies is the gradual transformation of the weak secondary (halogen) bonds into covalent bonds [1, 11, 12, 13, 16, 17]. This transformation is related by the appearance of hypercoordinated I atoms [1, 11, 12, 13, 16, 17], which in recent years have been associated with the existence of unconventional bonds named in the literature as metavalent bonds. Given the fact that high-pressure provides an efficient tool for e*ngineering* iodine *coordination*, the study of iodates under compression, is of crucial importance to deepen the knowledge of this unconventional bonding. Of particular interest is the study of iodate materials containing iodic acid (HIO$_3$). Such compounds combine different primary (covalent) and secondary (halogen) bonds, which might be tuned differently by pressure leading potentially to unexpected or non-classical behaviors which could test the current understanding of chemical bonding [18].

Recently, the syntheses of calcium, strontium and barium di-iodate hydrogen-iodate compounds were reported, namely, $Ca(IO_3)_2HIO_3$, $Sr(IO_3)_2HIO_3$ and $Ba(IO_3)_2HIO_3$ [19]. In these compounds, there is a coexistence of covalent and halogen bonds making them ideal candidates for high-pressure studies. In this work, we will focus on $Sr(IO_3)_2HIO_3$; a compound with a monoclinic structure described by the standard space group $P2_1/c$. We notice that in Ref. [19] the authors reported the non-standard space group $P2_1/n$, but this is a typo, the corresponding crystallographic information framework (CIF) file provided by the authors of Ref. [19], and uploaded to ICSD with code 140980, describes the $Sr(IO_3)_2HIO_3$ structure with space group $P2_1/c$, as we will continue to do so in this work. The crystal structure of $Sr(IO_3)_2HIO_3$ is represented in Figures 1a and 1b. In this monoclinic structure, there are 13 independent atoms all at *4e* Wyckoff sites: one Sr atom, three I atoms (labelled as I1 to I3), and nine O atoms (labelled as O1 to O9). For the naming of the atoms we use the same nomenclature used by An *et al*. [19]. The Sr atom is linked to eight oxygen atoms by ionic bonds: two from I1, two from I2, and four from I3 trigonal pyramids. The distorted $SrO_8$ polyhedron is asymmetric and two of these units are connected to each other via edge-sharing, thereby forming dimers of $Sr_2O_{14}$. Such dimers are interconnected via $IO_3$ pyramids forming layers that run nearly parallel to the *a-c* plane. These layers are separated by layers formed of $HOI_3$ and $IO_3$ units of I1.

Regarding the coordination of iodine atoms, the I1 and I3 atoms (i.e. the I atoms not related to the $HIO_3$ molecule,) are covalently bonded to three O atoms forming the classic trigonal pyramids of $IO_3$ (see Figure 1c). In the case of I3, the $IO_3$ pyramid is isolated and the three distances are nearly equal to 1.8 Å. In the case of I1, the $IO_3$ pyramid is close to the $HIO_3$ molecule and the three I-O bonds are nearly equal to 1.84 Å. A different bonding environment occurs for I2 (i.e. the I atom of the $HIO_3$ unit). There are two bonds close to 1.8 Å and a third one which corresponds to the elongated I–OH bond, with a distance of 1.94 Å. The presence of the hydrogen atom, at a distance of 0.85 Å from the oxygen involved in the 1.94 Å bond is what distorts the geometry of the $IO_3$ coordination pyramid of I2. In addition, for the I2 atom there is a fourth I2-O distance with a value of 2.28 Å. This distance corresponds to a bond, which is described in the literature as a weak secondary (halogen) I···O bond [20], between $HIO_3$ and the closest $IO_3$ pyramid (the one of the I1 atom), which can be classified as an intermolecular bond [21]. This bond is highlighted in Figure 1 using black color. As we show in this work, the distance of this longer I···O bond decreases rapidly under compression, favoring the formation of [HIO$_3$]·[IO$_3$] complexes. The next shortest weak secondary (halogen) I···O bond occurs for I1, since one oxygen of the IO$_3$ pyramid of I1 is at 2.47 Å from another IO$_3$ pyramid of the same kind. This bond is also highlighted in Figure 1 using grey color. It must be stressed that the direction of the two secondary bonds here mentioned corresponds to that of the iodine LEP. Within the donor-acceptor model for secondary bonding, this directionality has been associated with an electron density transfer from the LEP to the antibonding σ*(I-O) orbital from the closest IO$_3$ molecule [18]. However, the directionality can also be associated with the electrostatic interaction between the nucleophilic part (LEP) and the electrophilic part (σ* orbital) within the more recent σ-hole bonding model for secondary interactions, in which no charge transfer from the LEP to the σ* orbital is considered [22,23].

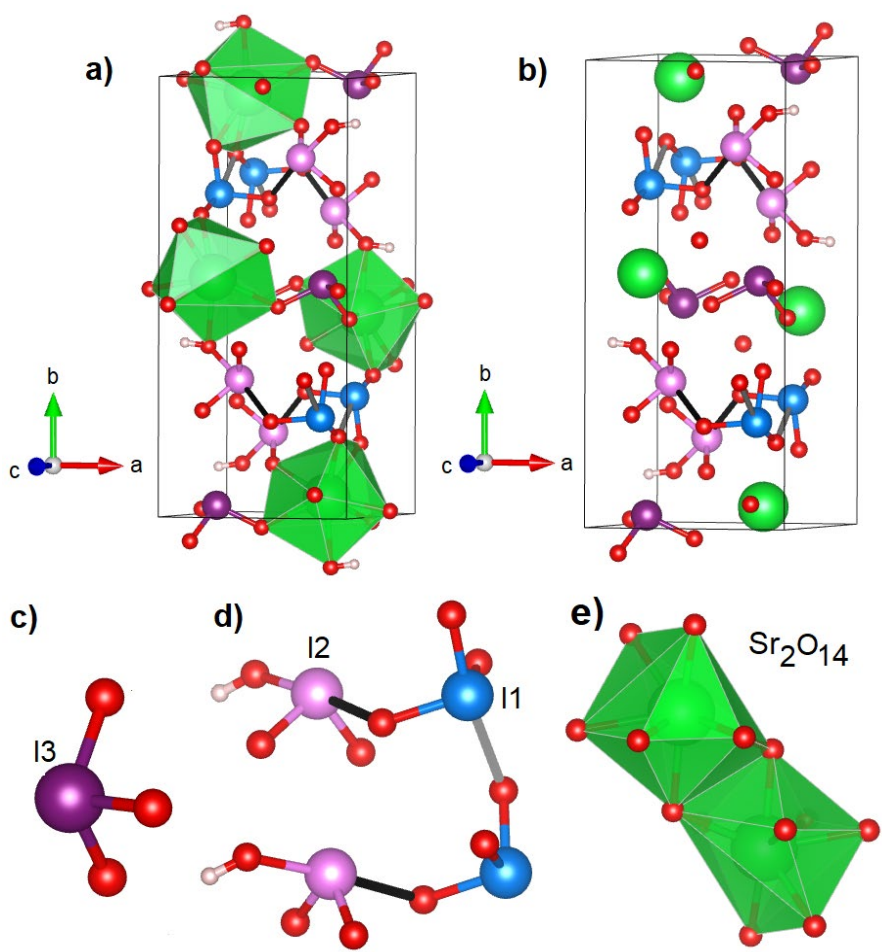

**Figure 1:** Crystal structure of Sr(IO$_3$)$_2$HIO$_3$. The I1, I2, and I3 atoms are shown in blue, magenta, and purple respectively. Oxygen, hydrogen, and strontium atoms are respectively shown in red, white, and green. (a) The structure shows the SrO$_8$ polyhedra in green and the coordination polyhedra of iodine atoms. (b) The structure only shows the coordination polyhedra of iodine atoms to facilitate the identification of their different bonds. The halogen I···O bond connecting HIO$_3$ to IO$_3$ is shown in black. The halogen I···O connecting IO$_3$ to IO$_3$ is shown in grey. c) The trigonal IO$_3$ coordination pyramid of I3. d) The [HIO$_3$]·[IO$_3$]·[IO$_3$]·[HIO$_3$] dimer. (e) The dimer of Sr$_2$O$_{14}$.

In this work, we study Sr(IO$_3$)$_2$HIO$_3$ under HP up to 20 GPa. We report the results of synchrotron powder X-ray diffraction (XRD), Raman spectroscopy, and optical-absorption experiments. The experiments are complemented by density-functional theory (DFT) calculations and the topology of the electron density has been analyzed within the Quantum Theory of Atoms in Molecules (QTAIM) [24]. We find that the Sr(IO$_3$)$_2$HIO$_3$ crystal structure can be described up to the highest pressure by space group *P*2$_1$/*c*, but pressure induces substantial changes within the layers of HOI$_3$ and IO$_3$ units, favoring iodine hypercoordination. Interestingly, the reported structural changes take place without the need for a reconstructive phase transition of first-order. In particular, the distance of the secondary (halogen) bonds linking the IO$_3$ molecules associated with the I1 and I2 atoms rapidly decreases under compression. This decrease of the secondary bond distance promotes first, the connection between [HIO$_3$] and [IO$_3$] units to form a [HIO$_3$]·[IO$_3$] complex, and second, then the connection of two [HIO$_3$]·[IO$_3$] complexes to form a [HIO$_3$]·[IO$_3$]·[IO$_3$]·[HIO$_3$] dimer.

A different behavior under pressure is found for the I3 atoms. Due to the isolation of the IO$_3$ pyramids associated with the I3 atoms from the rest of the IO$_3$ pyramids, they are not involved in the dimerization process. The different bond types found in Sr(IO$_3$)$_2$HIO$_3$ make the response of the crystal structure to pressure anisotropic. Other interesting phenomena we observed are the softening of several internal I-O vibrations of IO$_3$ units and a band-gap crossing. Under pressure, Sr(IO$_3$)$_2$HIO$_3$ changes from being a direct gap semiconductor to an indirect gap semiconductor. We also show that the decrease of the secondary (halogen) bonds under pressure leads to the formation of extra bonds with different characteristics. This phenomenon is associated with the softening of internal vibrations of IO$_3$ units [1, 11, 12, 13, 16]. The extra bonds are electron-deficient multicenter bonds (EDMB) giving account for the see-saw geometry around I1 and I2 atoms that occur at HP and that is schematized in Figure 1d.

## 2. Methods

### 2.1. Sample preparation

Single crystals of Sr(IO$_3$)$_2$HIO$_3$ were grown by a mild method in an aqueous solution at room temperature following the procedure described by Song *et al.* [26]. Sr(NO$_3$)$_2$ (5 mmol), HIO$_3$ (10 mmol), and K$_2$CO$_3$ (2.5 mmol) were mixed in a clean 50 ml beaker. Subsequently, 20 ml deionized water and 2 ml HNO$_3$ were consecutively added into the beaker under continuous stirring. After obtaining a homogenous mixture, the beaker was left in the open air to slowly evaporate for seven days. Colorless crystals, which were washed with distilled water, were obtained. The purity of the product and crystals was confirmed in a previous study by powder X-ray diffraction [26].

**2.2. High-pressure experiments**

All HP experiments were carried out using a membrane-type diamond-anvil cell (DAC) equipped with IIA-type diamonds with a culet size of 500 μm in diameter. We used 4:1 methanol ethanol as a pressure medium which remains quasi-hydrostatic up to 11 GPa [27]. The pressure was determined using ruby fluorescence with a precision of 0.05 GPa [28].

Powder XRD experiments were carried out at the BL04-MSPD beamline of the ALBA-CELLS synchrotron [29] using a monochromatic wavelength of 0.4642 Å and a spot size of the 20 μm × 20 μm at full-width-at-half-maximum. A Rayonix SX165 CCD image plate was used to collect the diffraction patterns. They were integrated into conventional XRD patterns using DIOPTAS [30]. FullProf [31] was used to perform Rietveld refinements and Le Bail fits.

HP Raman spectroscopy measurements were performed in single crystals using an inVia Renishaw Raman spectrometer system. We employed a 5× magnification objective, a laser wavelength of 632 nm, and a grating of 2400 lines·mm$^{-1}$. The set-up has a spectral resolution better than 2 cm$^{-1}$. The acquisition time for each spectrum was 20 s and the laser power was 10 mW before entering the diamonds of the DAC.

Optical absorption measurements were carried out on a crystal with parallel surfaces, a thickness of 10 μm, and faces of 60 μm x 60 μm. The absorbance was determined using the sample-in-sample-out method [32] using an optical set-up equipped with two confocal Cassegrain objectives and a USB4000-UV–VIS spectrometer from Ocean Optics.

**2.3. Computational details**

First-principles electronic structure calculations were performed under the formalism of DFT as implemented in the Vienna ab initio Simulation Package (VASP) [33 - 35]. We used the generalized gradient approximation (GGA) with the Perdew–Burke Ernzerhof (PBE) [36] and the PBE functional revised for solids (PBEsol) [37] parametrization for the exchange and correlation energy. The second parametrization has provided improved structural parameters that reasonably agree with experimental results. The optimization of the corresponding unit cell geometries and atomic positions was performed with the conjugate-gradient (CG) algorithm. The electron-ion interactions were represented by projected augmented-wave (PAW) pseudopotentials [38, 39]. The semi-core and valence states were treated explicitly by including the 1s electrons for H, 4s 4p 5s electrons for Sr, 2s 2p electrons for O, and 5s 5p electrons for I, with the remaining electrons being considered frozen at the core. The plane-wave kinetic-energy cutoff was defined as 600 eV, which together with the use of a dense Monkhorst-Pack grid [40] with a 6 × 3 × 6 k-point reciprocal space sampling mesh to ensure a convergence of the total energy of around 1 meV with deviations of the stress tensor from a diagonal hydrostatic form of less than 1 kbar (0.1 GPa). The vdW corrections to the total energy were applied within DFT by using the D3 method of Grimme with a zero-damping function [41].

Electronic band-structure calculations were carried out at different pressures along selected high-symmetry k-points on the first BZ. The three components of the dielectric function tensor were calculated using density perturbation theory applying the PBEsol and PAW methodology. In these calculations, a 9 x 4 x 8 Monkhorst-Pack grid was applied and tetrahedron smearing method applied on Brillouin zone. Once the dielectric tensor was calculated, we obtained the refractive index (n) given by $n = \sqrt{\frac{\sqrt{\varepsilon_1^2 + \varepsilon_2^2} + \varepsilon_1}{2}}$, where $\varepsilon_1$ and $\varepsilon_2$ are the real and the imaginary parts of the dielectric function. Phonon-dispersion curves were computed by using the supercell finite-displacement method implemented in the Phonopy package [42] with VASP as the force constant calculator. A 2 × 2 × 2 supercell was used to enable the exact calculation of frequencies at the zone center (Γ) and unique zone-boundary wave vectors. The phonon-dispersion curves were obtained on uniform 60 × 60 × 60 Γ-centered q-point meshes. The diagonalization of the dynamical matrix provides the normal-mode frequencies and allows the identification of the irreducible representations and the character of the vibrational phonon modes at the Γ-point. Crystal structures and related data were visualized and analyzed using VESTA [43]. The equation of state was determined using EoSFit [44].

The analysis of the electron density obtained from VASP calculations was performed with the CRITIC2 software [45] which integrates atomic properties in the framework of the QTAIM theory of Bader [24]. In particular, the Bader atomic charges at different pressures (Table S1 of the electronic supporting information (ESI)) as well as the renormalized number of electrons transferred (ET) between two atoms and the number of the electrons shared (ES) between two atoms (Table S2 of ESI) were obtained [24, 25, 46, 47]. The ES values of I-O, Sr-O, and H-O bonds is calculated as two times the delocalization index. The ET values of ionocovalent I-O and H-O bonds have been calculated as the charge of the I and H atoms divided by the nominal valence and also divided by the average number of O atoms to which the charge is transferred (3 for I and 1 for H) [25]. The ET values of ionic Sr-O bonds have been calculated dividing the Bader atomic charge for Sr by the nominal valence of Sr (2+).

## 3. Results and discussion

### 3.1 High-pressure effects on the crystal structure

In Figure 2 we present a selection of powder XRD patterns from $Sr(IO_3)_2HIO_3$. At room pressure (RP), Rietveld refinement confirms the monoclinic ($P2_1/c$) structure reported in the literature [19]. The goodness-of-fit parameters of the refinement of the patterns measured at RP are $R_p$ = 2.68% and $R_{wp}$ = 3.67%. We obtain unit-cell parameters of $a$ = 7.083(4) Å, $b$ = 15.943(8) Å, $c$ = 7.385(4) Å, and $β$ = 93.23(9)°, which agree with the literature [19]. At higher pressure, we do not notice the appearance or disappearance of any reflections. We only notice a change in the relative intensity of peaks due to preferred orientation effects and a broadening of peaks above 14.3 GPa due to the influence of non-hydrostatic stresses, which typically start to develop in the pressure medium beyond 11 GPa [21]. The two phenomena are typical of DAC experiments [48]. The development of preferred orientations is caused by a plastic deformation of the polycrystalline sample. The peak broadening is caused by the stress gradient in the radial direction, which is more evident under non-hydrostatic conditions. All XRD patterns can be indexed with the same space group ($P2_1/c$) as the structure at RP. Results of LeBail fits (including small residuals) are shown in Figure 2.

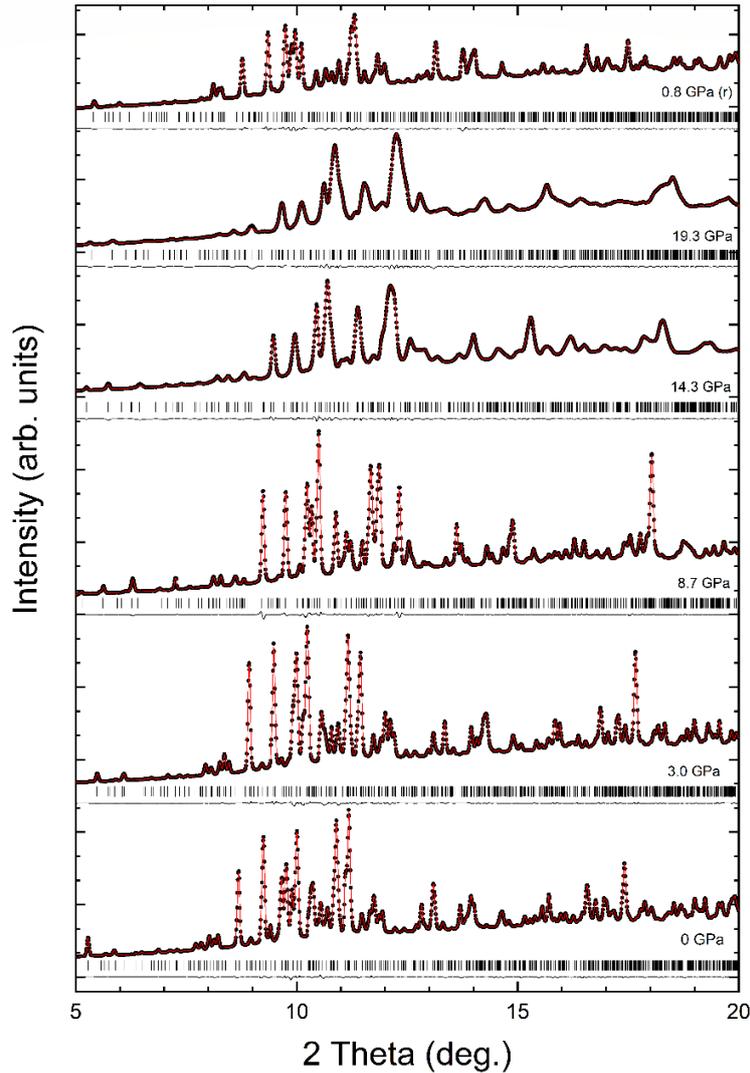

**Figure 2:** XRD patterns measured in Sr(IO₃)₂HIO₃ at different pressures and their corresponding fits. The experimental data are shown with black symbols, the fits with red lines, and the residuals with black lines. The vertical ticks indicate the positions of Bragg reflections. At 0 GPa, we show a Rietveld refinement; at other pressures, we show LeBail refinements. Pressures are indicated in the figure. 0.8 GPa (r) denotes a XRD pattern measured after pressure release.

Through the refinement of the powder XRD data, we obtained the pressure dependence of unit-cell parameters plotted in Figure 3. A good agreement between our experimental results and our DFT calculations is found. Up to around 3-4 GPa the *a*, *b*, and *c* parameters show similar compressibilities. At higher pressures, the *a* parameter becomes less compressible than the other two, and at around 20 GPa the *a* and *c* parameters become nearly identical. The change of compressiblity of the *a*-axis is connected to the formation of new bonds are dimmer units, that we will describe at the end of this section, which are mostly aligned along the direction of the *a*-axis [49]. The monoclinic β angle gradually increases under compression in the whole pressure range up to 20 GPa. From the pressure dependence of the volume, we conclude that Sr(IO$_3$)$_2$HIO$_3$ is very compressible. The volume decreases by approximately 18% from RP to 20 GPa with only a slight change in the slope of the pressure-volume curve at 5 GPa.

We have analyzed the pressure dependence of the volume using a third-order Birch-Murnaghan equation of state (EOS) [50] obtaining the volume at zero pressure ($V_0$), the bulk modulus at zero pressure ($K_0$), and its pressure derivative ($K_0'$), resulting in the following converged values: $V_0$ = 839(3) Å$^3$, $K_0$ = 23(2) GPa, and $K_0'$ = 6.7(6). These results are consistent with those obtained from calculations: $V_0$ = 832.8(6) Å$^3$, $K_0$ = 22(1) GPa, and $K_0'$ = 6.9(9). Therefore, we conclude that together with Zn(IO$_3$)$_2$ ($V_0$ = 265(1) Å$^3$, $K_0$ = 21.6(7) GPa, and $K_0'$ = 7.0(3)) [13] and Mg(IO$_3$)$_2$ ($V_0$ = 553(2) Å$^3$, $K_0$ = 22.2(8) GPa, and $K_0'$ = 4.2(4)) [16], Sr(IO$_3$)$_2$HIO$_3$ is one of the most compressible iodates studied to date.

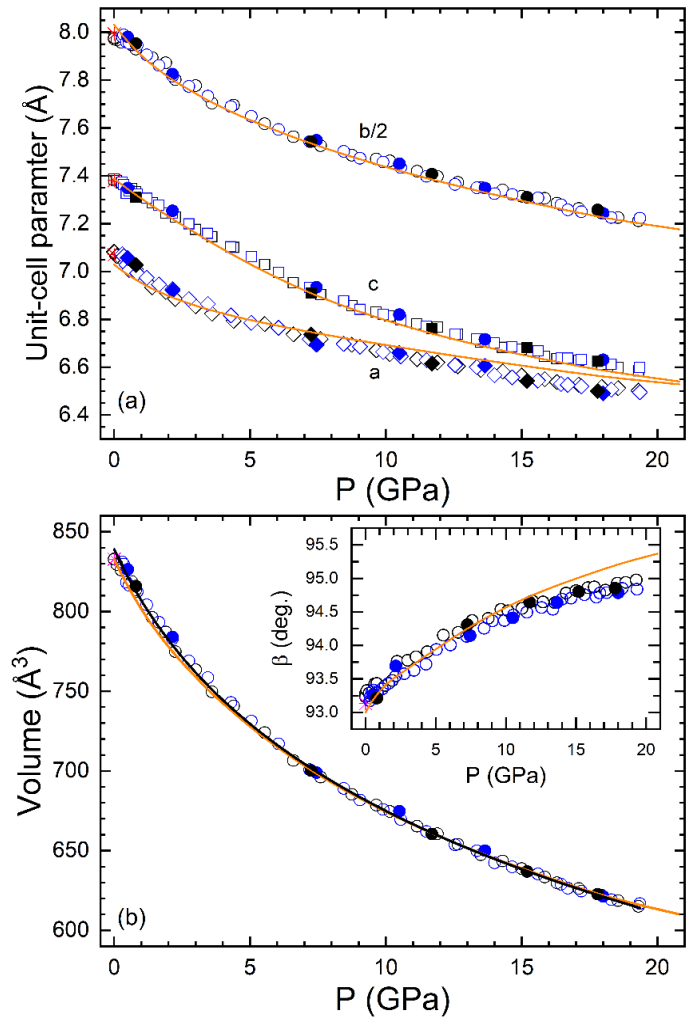

**Figure 3:** Pressure dependence of lattice parameters *a*, *b*, and *c* (a), and unit-cell volume (b). The inset in (b) shows the pressure dependence of the monoclinic β angle. Symbols are from experiments. Empty (solid) symbols are results obtained during compression (decompression). Blue and black colors are used to distinguish two different powder XRD experiments. The orange lines correspond to the results from DFT calculations. The black line in (b) is the EOS fit to the experimental data, as described in the text.

It must be noted that since the crystal structure of $Sr(IO_3)_2HIO_3$ is monoclinic, its compressibility must be properly described using the eigenvalues and eigenvectors of the non-diagonal isothermal compressibility tensor [51]. In our case, we have obtained them from experimental data using the PASCAL software [52]. The compressibility of the three main axes of compressibility are $9.7(2)\times10^{-3}$, $9.4(3)\times10^{-3}$, and $5.9(2)\times10^{-3}$ $GPa^{-1}$. The most compressible direction is [304], which is the direction of the weak intermolecular secondary I⋯O bonds between the $IO_3$ pyramids of I1 and I2. The second most compressible direction is [010]; i.e. the direction perpendicular to the layers formed by $HIO_3$ and $IO_3$ units. The least compressible direction is [40$\bar{3}$], which coincides with the orientation of the $Sr_2O_{14}$ dimers.

To better understand the changes induced by pressure in the crystal structure it is important to analyze the changes induced by pressure in the first- and second-neighbor bond distances. This information cannot be obtained from these powder XRD experiments due to the influence of preferred crystallite grain orientations which prevents successful Rietveld refinement. However, given the excellent agreement between experiments and calculations regarding the pressure dependence of unit-cell parameters, analysis of the pressure-evolution of bond distances can be performed using the bond distances provided by our DFT calculations. As can be observed in Figure 4, the simulated bond lengths at RP are comparable to those of the Rietveld refinement performed on the RP diffraction data.

To discuss bond lengths, we will use the bond classification proposed by Jansen and coworkers [53] for iodates, but reinterpreted according to a recent work of Manjón and coworkers [25]. According to Ref. [53], the strong and short I–O bonds (distances < 2.2 Å) can be classified as primary (covalent) bonds, while the weak and long I⋯O bonds (distances from 2.2 to 2.6 Å) can be classified as secondary (halogen) intermolecular bonds. According to Ref. [25], this classification has to be slightly modified since the covalent I-O bonds should have bond lengths of the order of 1.8 Å, and the I-O bonds with an intermediate bond length (between 1.9 and 2.2 Å) form a third type of bond with an intermediate strength. This bond, previously called as metavalent bond in iodates [1, 11, 12, 13, 16], due to its similarity to bonds formed in hypercoordinated units at HP in some IV-VI and $V_2VI_3$ chalcogenides [46, 47, 54, 55], is in fact an electron deficient multicenter bond (EDMB) similar to those which occur in diborane and in the octahedrally-coordinated phases of pnictogens and chalcogens [25]. According to this picture, it can be considered that the I1 and I3 atoms in $Sr(IO_3)_2HIO_3$ form covalent bonds at RP with their three first-neighbor oxygen atoms (at distances around 1.8 Å) forming trigonal pyramids as described in the introduction and illustrated in Figure 1. As aforementioned, the $IO_3$ pyramid of I3 is isolated from those of the I1 and I2 atoms. Consequently, for the I3 atom, the second-neighbor oxygen atoms are at distances larger than 2.95 Å; however, I1 forms also a secondary (halogen) I⋯O bond with a second-neighbor oxygen atom from the neighboring $IO_3$ pyramid formed by another I1 atom that is at 2.47 Å (as shown by the grey bond in Figure 1d). The next closest oxygen atoms to I1 are beyond 2.66 Å. Additionally, the I2 atom forms two covalent bonds (distances around 1.8 Å) and one EDMB formed by two quasi-aligned I-O bonds. In this O-I-O EDMB, the shortest I-O bond has a length of 1.94 Å and the longest I-O bond has a length of 2.28 Å (black bond in Figure 1d). The next closest oxygen atoms to I2 are at a distance larger than 3.07 Å.

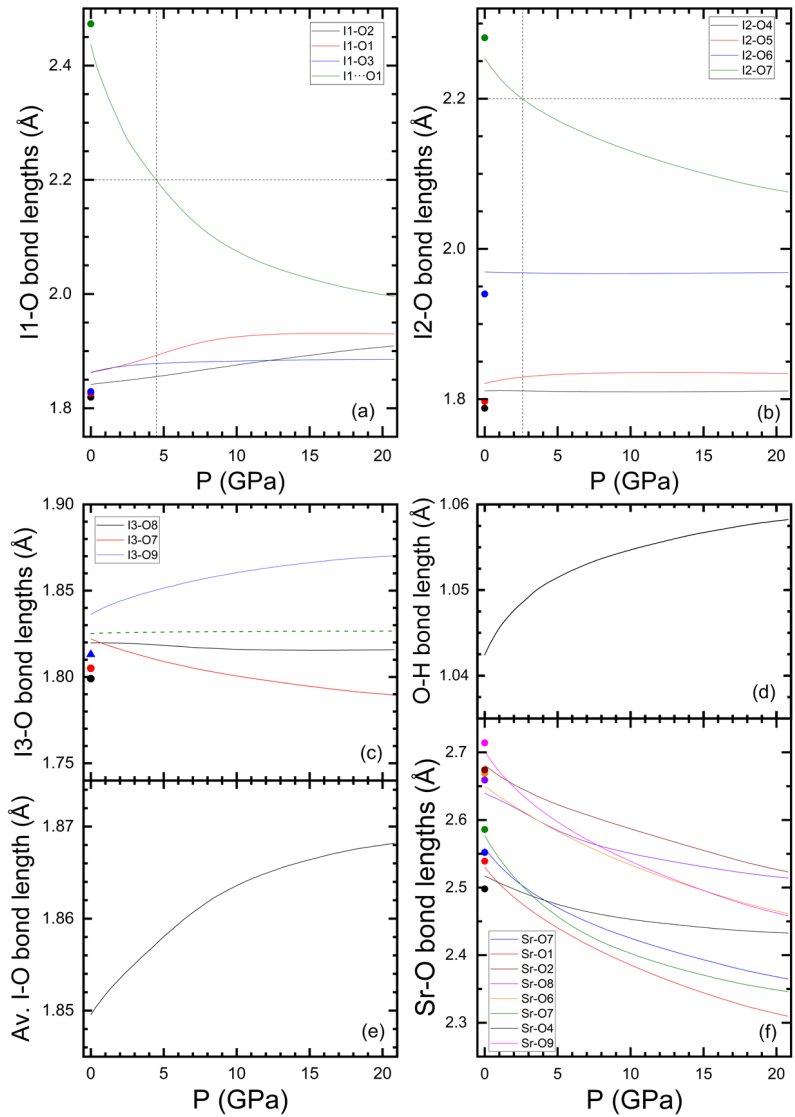

**Figure 4:** Pressure dependence of I-O, O-H, and Sr-O bond distances in $Sr(IO_3)_2HIO_3$. Symbols are results from experiments at room pressure and lines are results from calculations. (a) shows the I-O bond length for I1, (c) for I2, (c) for I3, (d) shows the O-H bond length, (e) shows the average of the calculated I-O distances for internal bonds of the three $IO_3$ units, and (f) shows the Sr-O bond lengths. In (a) and (b) the dotted lines show the pressure where the covalent bond is formed. In (c) the dashed line is the average I-O bond distance for I3. The colors used for individual bonds are shown in the legends. I1⋯O1 is the halogen bond.

Under compression, the different bonds described in the previous paragraph exhibit different behaviors. For example, the O-H bond distance increases slightly under compression (see Figure 4e), whereas the Sr-O bond distances decrease (see Figure 4f) in such a way that the SrO$_8$ polyhedron becomes slightly more distorted. The compressibility of the Sr-O bonds is similar to that reported for the same bond in SrWO$_4$ [56] and SrCrO$_4$ [57]. The monotonous behavior of O-H and Sr-O bonds under compression contrasts with the changes observed in the I-O bond distances. In Figure 4c it can be seen that for I3 (the I atom forming the isolated IO$_3$ units) one I3-O bond is not affected by pressure, one shows a slight decrease with pressure, and the third one shows a slight increase with pressure. The average value of the I3-O bond distance (the dashed line in the plot) remains nearly constant up to 20 GPa, which suggests that all I3-O bonds remain as covalent bonds in this pressure range. This result is explained by the isolation of the IO$_3$ unit of I3 previously mentioned (the next second-neighbor distance is well above 2.5 Å for P < 20 GPa) and contrasts with the increase under compression of the average bond length of the three shortest I-O bonds when the three IO$_3$ units of I1, I2, and I3 are considered (see Figure 4e). The fact that the shortest I-O bonds increase in length under compression is expected to considerably reduce the bandgap [14], a phenomenon that we indeed observe and that will be discussed in section 3.3.

A completely different behavior is observed in Figure 4a for I1, the other I atom not related to the HIO$_3$ molecule. It can be seen that the three short distances of the IO$_3$ pyramid are enlarged under compression, with the short I1-O1 distance at 0 GPa being the most elongated (see Figure 4a). In contrast, the long secondary I1⋯O1 bond rapidly decreases with pressure, becoming 2.2 Å at 4.5 GPa. The decrease of the I1⋯O1 distance leads to the formation of a new I1-O1 bond that keeps decreasing in length after its formation but with a smaller compressibility, reaching a value of ca. 2.0 Å at 20 GPa. The new bond leads to the fourfold coordination (hypercoordination) of the I1 atom and the new IO$_4$ polyhedron which has a distorted see-saw geometry with two long bonds (I1-O1) and two short bonds (I1-O2 and I1-O3).

According to the interpretation of Jansen *et al.* [53], the halogen I1⋯O1 bond at RP transforms into a covalent I-O bond at HP (when the I-O bond distance goes below 2.2 Å) forming a IO$_4$ unit (see Figure 1d), where the iodine atom is hypercoordinated. Note that the see-saw geometry of the IO$_4$ unit corresponds to a trigonal bipyramidal configuration when the LEP of iodine, which is in the same plane as the I1-O2 and I1-O3 bonds and is almost perpendicular to the direction of the two I1-O1 bonds, is considered. This result can be interpreted using the bonding model proposed by Manjón *et al.* [25], according to which the decrease of the I1⋯O1 distance leads to the formation of a new I1-O1 bond with a length between 2.2 and 2.0 Å between 4.5 and 20 GPa (see Figure 4a) that is part of a quasi-linear O1-I1-O1 three-center-two-electron (3c-2e) EDMB with the O1-I1-O1 angle being 167.43° at RP. The 3c-2e EDMB is formed based on the basis of the short I1-O1 bond since the short I1-O1 bond length increases from 1.86 to 1.93 Å between 0 and 20 GPa (see Figure 4a). Therefore, the I1 atom becomes hypercoordinated with two short I-O bonds and two long I-O bonds forming a see-saw hypercoordinated multicenter unit [25]. The pressure-induced formation of the fourth I-O bond around the I1 atom and the changes in the bond distances of the original IO$_3$ unit can be interpreted in the following way. As the I1···O1 distance decreases to form an extra I1-O1 bond, there is a charge transfer to the secondary (halogen) I1···O1 bond from the covalent I1-O1 bond that is almost opposite to it. The charge transfer is responsible for the progressive lengthening of the short I1-O1 bond. Such charge transfer weakens the short I1-O1 covalent bond and hardens the newly formed I1-O1 bonds, as has already been demonstrated in HIO$_3$ under pressure [16]. This charge transfer model from the primary bond to the secondary bond is clearly different from the donor-acceptor model [18] and more in line with the σ-hole model for secondary bonding [22,23].

To demonstrate the pressure-induced formation of the O1-I1-O1 3c-2e bond and the charge transfer from the short covalent I1-O1 bond towards the secondary I1···O1 bond, we have plotted in Figure 5a the pressure dependence of the ES values of the shortest I1-O bonds. As can be observed, the ES values of the covalent I1-O2 and I1-O3 bonds slightly decrease at HP. In contrast, the ES values of the short and long I1-O1 bonds not only show very large changes with pressure but also an inverse behavior that confirms the charge transfer from the short I1-O1 bond towards the long I1-O1 bond as pressure increases. Both bonds tend to have almost the same ES value at 20 GPa. This charge transfer shows the multicenter character of the newly formed O1-I1-O1 3c-2e EDMB [25]. Moreover, it explains why the strong short covalent I1-O1 bond decreases in strength while the weak secondary halogen I1···O1 bond increases in strength as pressure increases. This change in the strength of the two opposite and correlated bonds also occurs in HIO$_3$ when the iodine atom becomes hypercoordinated under compression [17]; a fact that can be explained by the formation of EDMBs.

Regarding I2 (the I atom of the HIO$_3$ unit), Figure 1d shows that I2 exhibits almost a fourfold coordination at RP. It is linked to oxygen atoms at 1.797, 1.788, and 1.940 Å within HIO$_3$, and has another bond with a length of 2.281 Å to an oxygen atom of the IO$_3$ pyramid of I1. Thus, it can be considered that I2 almost forms a quasi-linear O3-I2-O6 3c-2e EDMB at RP. Indeed, we can interpret that the two short I2-O4 and I2-O5 bonds of similar lengths (below 1.80 Å at RP) correspond to covalent bonds, while the third short I2-O6 bond, with length 1.940 Å, is not a pure covalent bond. The reason is that there is also a charge transfer from the I2-O6 bond to the secondary I2···O3 bond, with a length of 2.281 Å, which is almost opposite to the I2-O6 bond. Note that the secondary I2···O3 bond is considerably shorter than the previously mentioned secondary I1···O1 bond, so the charge transfer is stronger at RP for the I2-O6 bond than for the short I1-O1 bond. Figure 4b shows that the three I-O bonds within the $HIO_3$ unit barely change with pressure despite the fact that there is a considerable decrease of the I2···O3 bond distance that goes below 2.2 Å at very low pressures. This behavior can be understood if we consider that the hypercoordinated multicenter see-saw geometry of the I2 atom is already formed at RP.

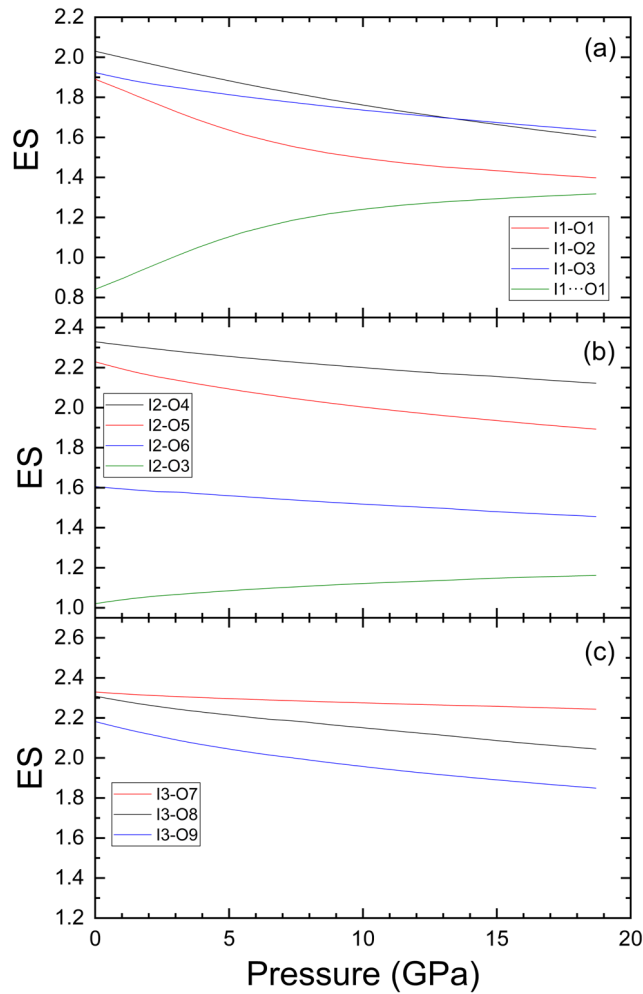

**Figure 5:** Pressure dependence of the ES values of different I-O bonds in $Sr(IO_3)_2HIO_3$. (a) I1-O bonds (b) I2-O bonds, and (c) I3-O bonds. We have used the same color code as in Figure 4.

As already made for the I1 atom, the charge transfer from the I2-O6 bond to the I2-O3 bond upon the formation of the O3-I2-O6 3c-2e EDMB can be traced by the evolution of the ES values of both bonds as a function of pressure (Figure 5b). The ES values of I2-O3 and I2-O6 bonds show an inverse behavior with pressure that tends towards a common value. This result reaffirms that there is a charge transfer from the I2-O6 bond towards the extra I2-O3 bond as pressure increases which is consistent with the elongation and weakening of the I2-O6 bond at HP. The behavior of ES for the I2-O6 bond at HP contrasts with those of the covalent I2-O4 and I2-O5 bonds, which are similar to those of covalent I1-O2 and I1-O3 bonds (Figure 5a) and also similar to the three short covalent I3-O bonds (Figure 5c).

As a consequence of the changes triggered by pressure in the coordination environment of I1 and I2 described above, a [HIO$_3$]·[IO$_3$]·[IO$_3$]·[HIO$_3$] dimer is formed (see Figure 1d). In the first step, it could be considered that a [HIO$_3$]·[IO$_3$] complex is formed at 2.6 GPa and subsequently, the dimer is formed at 4.5 GPa. Interestingly, this pressure coincides with the pressure where we observed a change in the compressibility of the unit-cell parameters (see Figure 3a). The formation of this supramolecular network indicates a decrease of the anion⋯anion electrostatic repulsion as pressure increases and a partial suppression of the stereochemical activity of the LEP of iodine [18]. As will be shown in section 3.3, the formation of the dimer has consequences in the electronic band structure, transforming Sr(IO$_3$)$_2$HIO$_3$ from a direct wide-gap semiconductor into an indirect wide-gap semiconductor. On the other hand, the enlargement of the short I1-O1 bond with pressure drives a softening of internal vibrational modes of IO$_3$ as we will show in section 3.2. Finally, it must be noted that the see-saw geometry of the IO$_4$ units in the dimer shown in Figure 1d is similar to that formed by the TeO$_4$ units in tetragonal paratellurite ($\alpha$-TeO$_2$) [58] and also similar but slightly different to that formed by the TeO$_4$ units in orthorhombic tellurite ($\beta$-TeO$_2$) [59].

**3.2 High-pressure effects on lattice vibrations**

The primitive cell of Sr(IO$_3$)$_2$HIO$_3$ contains four formula units and therefore 168 vibrations are expected. According to group-theory analysis, the mechanical representation of the vibrations is $\Gamma = 42A_g + 42A_u + 42B_g + 42B_u$. Even (*gerade*) modes ($A_g$ and $B_g$) are Raman active. Three of the odd (*ungerade*) modes ($A_u + 2B_u$) correspond to acoustic modes. The rest of the $A_u$ and $B_u$ modes are IR active. In Table S3 of ESI, we report the calculated frequency of the 165 optical modes of Sr(IO$_3$)$_2$HIO$_3$ including their assignment. In Figure 6 we report the Raman spectra of a crystal of Sr(IO$_3$)$_2$HIO$_3$ at ambient conditions outside the DAC, where 36 modes have been identified. The measured frequencies are summarized in Table 1 and compared with a previous study [19] and present DFT calculations. The agreement between experiments and calculations is excellent.

The Raman and IR spectra of Sr(IO$_3$)$_2$HIO$_3$ at ambient conditions have been previously reported by An *et al.* [19]. These authors reported only fifteen Raman modes and five absorption bands in the IR spectrum. Curiously, one IR absorption band observed at 1574 cm$^{-1}$ in Sr(IO$_3$)$_2$HIO$_3$ has been also observed at a similar frequency in Ca(IO$_3$)$_2$HIO$_3$ despite the fact that no phonons are expected for Sr(IO$_3$)$_2$HIO$_3$ between 1200 to 2450 cm$^{-1}$ according to our DFT simulations. The frequency of the IR absorption band might be correlated to one of the absorption bands of HNO$_3$ [60], one of the precursors of the synthesis of both Sr(IO$_3$)$_2$HIO$_3$ and Ca(IO$_3$)$_2$HIO$_3$. We consider the IR absorption at 1574 cm$^{-1}$ reported by An *et al.* [19] to be likely due to the presence of a small fraction of HNO$_3$ in their crystals. According to our simulations, the IR modes around 1200 and 2470 cm$^{-1}$ could also be assigned to O–H vibrations. The vibrational modes in Sr(IO$_3$)$_2$HIO$_3$ can be separated between the external and internal (stretching and bending) modes of the iodate anions. Modes with frequencies above 550 cm$^{-1}$ can be assigned to I–O stretching vibrations of different IO$_3$ pyramids. In particular, the strongest mode of the Raman spectrum, at 758.8 cm$^{-1}$, which is observed at similar frequencies in most iodates [1, 11, 12, 13, 16], is associated to a symmetric stretching I-O vibration. On the other hand, the modes from 200 to 400 cm$^{-1}$ are related to bending I-O vibrations. The lowest frequency modes (below 200 cm$^{-1}$) involve external vibrations of the iodate anions and of the Sr atoms.

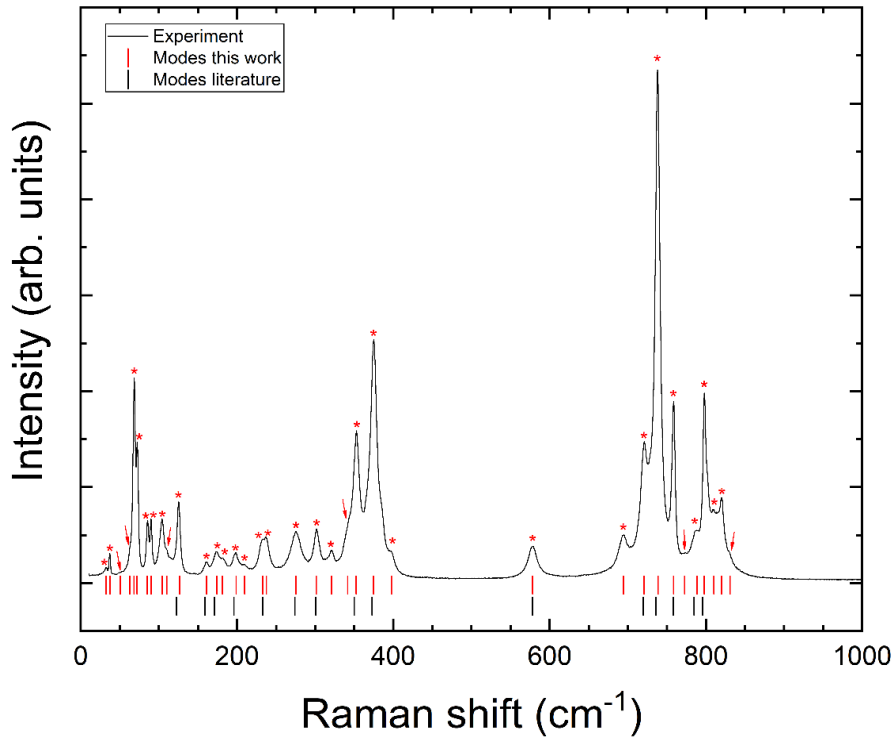

**Figure 6:** Raman spectrum of Sr(IO$_3$)$_2$HIO$_3$ measured at ambient conditions. The vertical ticks indicate the frequency of the modes from this work (red) and the previous work (black). The modes are also identified by asterisks or arrows in the figure.

In Figure 7 we present a selection of Raman spectra measured at different pressures. In the Raman experiments we did not find any evidence of a structural phase transition. The most noticeable changes induced by pressure in the Raman spectra are the shift of low-frequency modes, especially the bending modes, towards higher frequencies and the expansion of the frequency region covered by stretching (high-frequency) modes. This last feature is due to the gradual softening with pressure of the stretching modes with the smallest frequencies and the hardening with pressure of the stretching modes with the highest frequencies. The softening of the stretching modes with the smallest frequencies together with the hardening of the bending modes also causes the closing of the phonon gap between both types of modes as pressure increases. The different pressure dependence of the Raman modes changes the appearance of the Raman spectrum change, however, no appearance/disappearance of modes is identified. Because of different pressure dependencies, several modes overlap under compression. At the lowest pressure measured with the DAC (0.57 GPa), we observe 35 modes which is consistent with Raman spectra acquired at ambient conditions. At the highest pressure we observe only 28 modes.

| | Raman | | | | Infrared | |
|---|---|---|---|---|---|---|
| Mode | Experiment This work $\omega$ (cm$^{-1}$) | DFT This work $\omega$ (cm$^{-1}$) | Experiment Literature $\omega$ (cm$^{-1}$) | Mode | DFT This work $\omega$ (cm$^{-1}$) | Experiment Literature $\omega$ (cm$^{-1}$) |
| $A_g$ | 32.6 | 36.9 | | $A_u$ | 697.9 | 699 |
| $A_g$ | 37.4 | 41.5 | | $B_u$ | 773.7 | 766 |
| $B_g$ | 50.5 | 58.6 | | $A_u$ | 1206.6 | 1151 |
| $B_g$ | 62.8 | 66.7 | | | | 1574 |
| $A_g$ | 68.4 | 74.3 | | $B_u$ | 2473.6 | 2707 |
| $B_g$ | 72.3 | 77.6 | | | | |
| $B_g$ | 85.0 | 84.0 | | | | |
| $A_g$ | 90.0 | 90.8 | | | | |
| $B_g$ | 104.2 | 103.7 | | | | |
| $B_g$ | 110.1 | 111.8 | | | | |
| $A_g$ | 124.6 | 123.2 | 123 | | | |
| $A_g$ | 159.0 | 152.3 | | | | |
| $B_g$ | 161.0 | 153.7 | 159 | | | |
| $A_g$ | 174.4 | 180.0 | 171 | | | |
| $B_g$ | 181.4 | 183.7 | | | | |
| $A_g$ | 199.0 | 192.7 | 196 | | | |
| $B_g$ | 210.0 | 209.1 | | | | |
| $B_g$ | 232.8 | 233.3 | 233 | | | |
| $A_g$ | 237.6 | 239.1 | | | | |
| $A_g$ | 275.2 | 283.3 | 274 | | | |
| $A_g$ | 301.4 | 306.6 | 301 | | | |
| $A_g$ | 321.0 | 324.9 | | | | |
| $B_g$ | 341.7 | 339.6 | | | | |
| $B_g$ | 352.8 | 352.6 | 350 | | | |
| $B_g$ | 374.8 | 382.6 | 373 | | | |
| $A_g$ | 397.6 | 395.2 | | | | |
| $B_g$ | 578.4 | 566.6 | 578 | | | |
| $A_g$ | 694.4 | 693.4 | | | | |
| $B_g$ | 720.9 | 715.5 | 720 | | | |
| $A_g$ | 736.7 | 731.7 | 736 | | | |
| $B_g$ | 758.8 | 742.6 | | | | |
| $A_g$ | 773.0 | 749.2 | | | | |
| $B_g$ | 788.6 | 761.1 | | | | |
| $A_g$ | 798.0 | 779.9 | | | | |
| $B_g$ | 810.3 | 782.3 | | | | |
| $A_g$ | 820.1 | 784.8 | | | | |
| $A_g$ | 826.8 | 785.8 | | | | |

**Table 1:** Calculated and measured Raman and IR frequencies in Sr(IO$_3$)$_2$HIO$_3$ at room conditions. Those previously reported in the literature [19] are also shown for comparison.

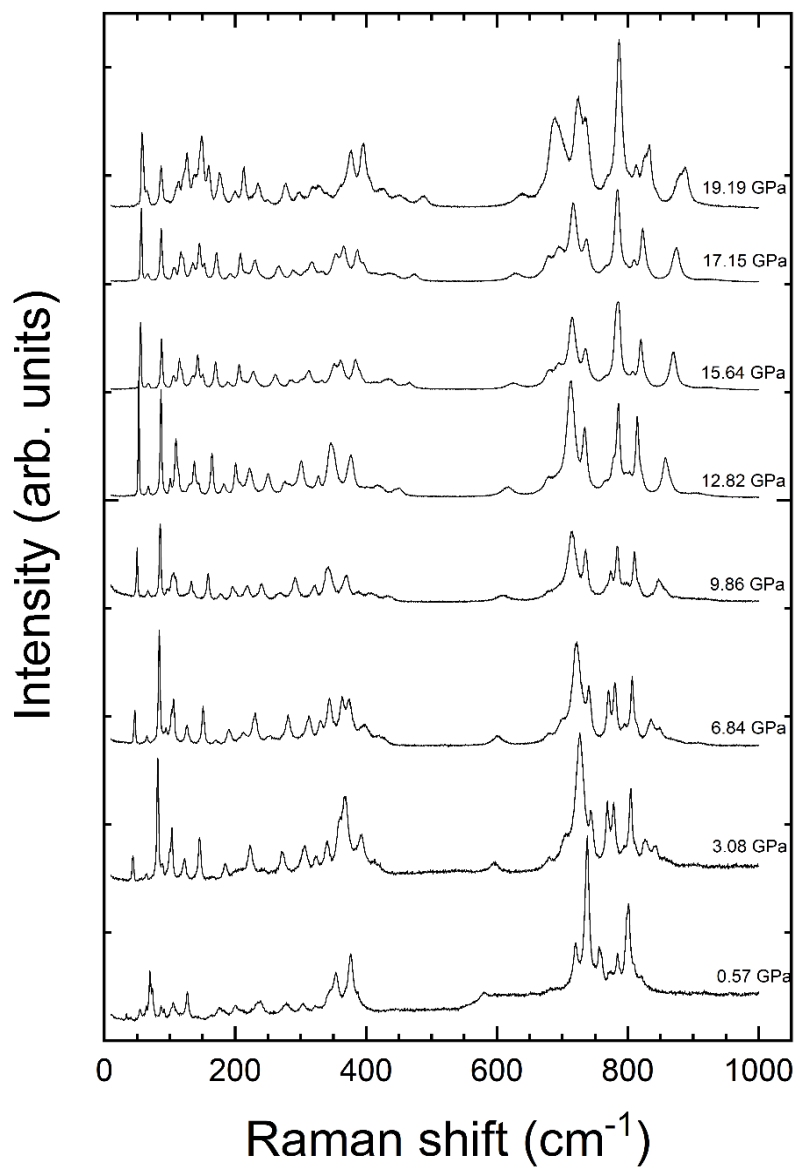

**Figure 7:** Selection of Raman spectra of Sr(IO$_3$)$_2$HIO$_3$ measured at different pressures.

From the experimental Raman spectra, we have obtained the pressure dependence of the Raman modes in Sr(IO$_3$)$_2$HIO$_3$. In this task, the help of calculations was crucial to identify the different Raman modes because, as pressure was increased, several phonon crossings took place. The pressure dependence of the observed Raman modes is reported in Figure 8. In Fig. 8a we show the pressure dependence of the low-frequency modes. Most of them follow a non-linear behavior with slope changes between 3 and 5 GPa. Quadratic fits describing the pressure dependence of modes are reported in Table S4 of ESI. In particular, there is a low-frequency mode, represented by open black symbols (experiments) and a black dashed line (calculations) in which the sign of the slope changes from positive to negative. In the figure, it can be also seen several phonon crossings between $A_g$ and $B_g$ modes. In Fig. 8b we show the pressure dependence of the high-frequency modes. It can be seen that the bending mode at 578 cm$^{-1}$ is one of the modes that most rapidly increases under pressure. Similarly, all the high-frequency stretching modes (above 750 cm$^{-1}$) rapidly shift towards high frequency. In contrast, we have observed that all the low-frequency stretching modes (between 690 and 750 cm$^{-1}$) undergo a frequency decrease as pressure increases below 5 GPa. This is quite evident for the strongest mode (an I-O stretching vibration) that shifts quickly towards low frequencies below 5 GPa. However, once the dimerization occurs above 5 GPa, a change occurs in the pressure coefficients of all these soft modes and their frequencies shift slowly towards high frequencies. The changes in the pressure dependence of these Raman modes around 5 GPa are consistent with the hypercoordination and dimer formation that gradually happens up to this pressure, which was described in the previous section. In addition, the softening observed in some I-O stretching modes correlates well with the enlargement under pressure of several I-O bond lengths commented on in the previous section and also in line with results from previous works on iodates [11, 12, 13, 16].

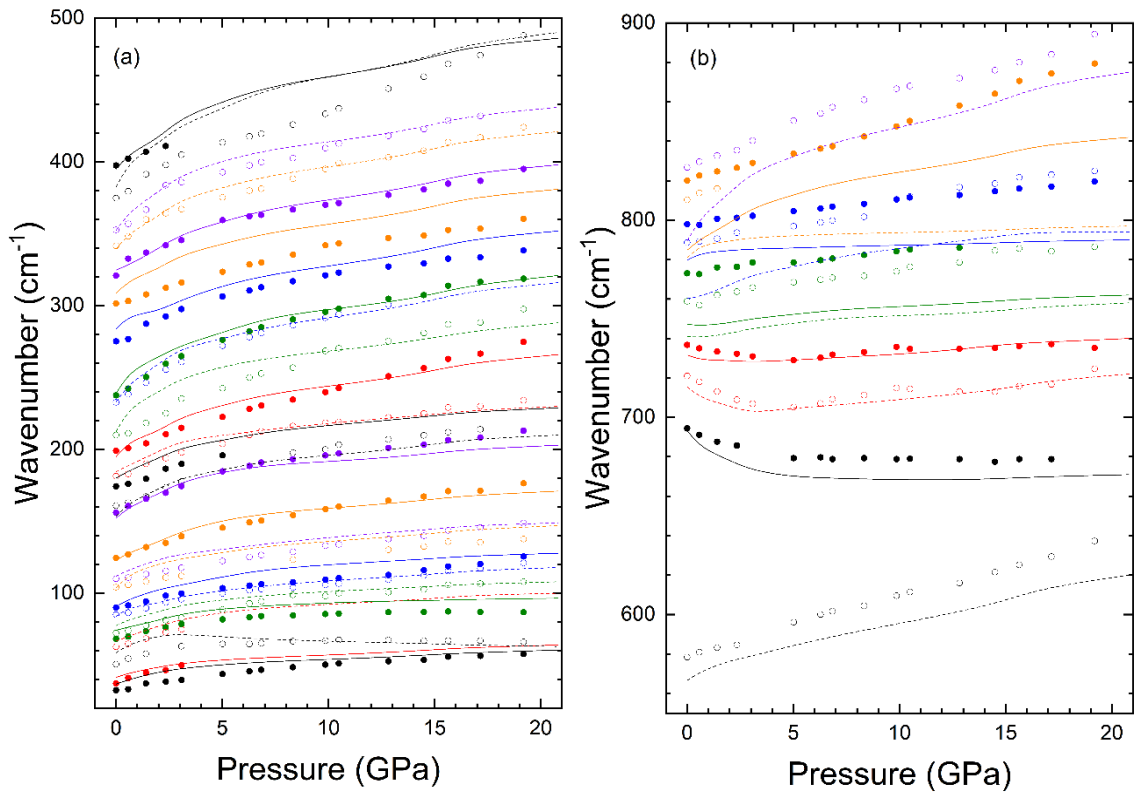

**Figure 8:** Pressure dependence of the Raman mode frequencies in Sr(IO$_3$)$_2$HIO$_3$: (a) low-frequency region, and (b) high-frequency region. We use solid (empty) symbols for $A_g$ ($B_g$) modes measured in experiments. We use solid (dashed) lines for calculated $A_g$ ($B_g$) modes. Different colors are only used to facilitate the identification of modes in the figure. Note that only the theoretical modes tentatively assigned to the experimental ones have been plotted for the sake of clarity. All theoretical modes are given in the Supporting Information.

Taking into account the above considerations, the phonon behavior of iodates under pressure can be explained in the following way. Since there is an inverse relationship between the frequencies of stretching modes and the bond distance, the highest frequency modes in iodates (typically above 750 cm$^{-1}$) correspond to the shortest I-O bonds, which are the covalent I-O bonds (with bond lengths around 1.8 Å). As we have observed in Sr(IO$_3$)$_2$HIO$_3$, the I-O covalent bond lengths are not significantly pressure-dependent, so the decrease of the crystal volume leads to an increase in the charge density of covalent I-O bonds which explains the hardening of their related high-frequency stretching modes with pressure. The hardening of the high-frequency stretching modes in Sr(IO$_3$)$_2$HIO$_3$ has not been observed in other iodates, such as Co(IO$_3$)$_2$ [11], Fe(IO$_3$)$_3$ [12], Zn(IO$_3$)$_2$ [13], and Mg(IO$_3$)$_2$ [16], because in these iodates the I-O bond lengths increase with increasing pressure. Thereby causing a softening of the stretching modes. The softening of stretching modes is also observed in Sr(IO$_3$)$_2$HIO$_3$ but it only affects the lowest frequency stretching modes (below 750 cm$^{-1}$) as already commented. These modes are related to the largest I-O bonds that correspond to the less pure covalent I-O bonds or even the I-O bonds related to the O-I-O 3c-2e EDMBs (with bond lengths between 1.9 and 2.3 Å in Sr(IO$_3$)$_2$HIO$_3$). These less pure covalent bonds (all I-O bonds in the mentioned iodates [11, 12, 13, 16] as well as the short I1-O1 and short I2-O6 bonds in Sr(IO$_3$)$_2$HIO$_3$) tend to increase their lengths with increasing pressure at low pressures; a fact related to its loss of charge since it is transferred to the extra bonds formed (the relatively long I1-O1 and I2-O3 bonds in Sr(IO$_3$)$_2$HIO$_3$).

Unlike in Sr(IO$_3$)$_2$HIO$_3$, the process of new bond formation is not completed at the highest pressure reached in the other mentioned iodates (Co(IO$_3$)$_2$ [11], Fe(IO$_3$)$_3$ [12], Zn(IO$_3$)$_2$ [13], and Mg(IO$_3$)$_2$[16]). Both the increase of I-O bond length and the decrease of the electronic charge as pressure increases result in a pressure-induced bond softening that causes the decrease of phonon frequency with increasing pressure observed in these low-frequency stretching modes. Once the extra bonds are formed and the charge transfer ends (above 5 GPa in Sr(IO$_3$)$_2$HIO$_3$), the bond lengths of the long I1-O1 and I2-O6 bonds cease to increase. Concomitantly, the decrease of the crystal volume with pressure leads to a normal increase of the charge density of the formed EDMB as well as of the covalent I-O bonds. These changes result in a different pressure behavior of the soft modes that start to harden similarly as the high-frequency stretching modes related to the unperturbed covalent I-O bonds.

Finally, it must be noted that the bending modes in iodates (typically between 200 and 600 cm$^{-1}$) can be considered to be optical modes that derive from acoustic modes at the Brillouin zone edge. Note that the acoustic modes can become optical modes due to the folding of the Brillouin zone edge into the Brillouin zone center as the crystal symmetry is reduced. Therefore, the strong hardening of the bending modes in iodates, in particular the highest frequency bending modes, is related to the strong hardening of the acoustic modes in materials undergoing the formation of extended three-center bonds.

**3.3 High-pressure effects on the electronic properties**

In Figure 9 we show a selection of absorption spectra measured in $Sr(IO_3)_2HIO_3$ at selected pressures. The absorbance has a sharp absorption edge at high energy, which corresponds to the fundamental band gap, plus an exponential sub-band gap absorption, which resembles an Urbach tail and is typical of iodates [14] and other oxides [61]. The shape of the absorption edge does not allow the identification of the band gap as direct or indirect. According to our band structure calculations, $Sr(IO_3)_2HIO_3$ is a direct gap material at RP. However, the difference between the direct gap and the closest indirect gap is smaller than 0.1 eV. As we will discuss next, under compression the situation reverses, but the difference in energy between the two gaps is always smaller than 0.1 eV. Usually for the direct gap the absorption coefficient is stronger than the absorption coefficient for an indirect gap [62] and if the two gaps are close in energy, as in $Sr(IO_3)_2HIO_3$, the absorbance as a first approximation can be interpreted as that of a direct gap to determine the band-gap energy [63]. Under this assumption the band-gap energy was determined by plotting the square of the absorption coefficient against the photon energy and extrapolating the linear region of it to the abscissa of the plot (see inset of Fig. 9). We have determined a band-gap energy of 4.14(5) eV at 0.28 GPa in agreement with the previous values reported at RP, 4.2 eV [19] and 4.1 eV [26].

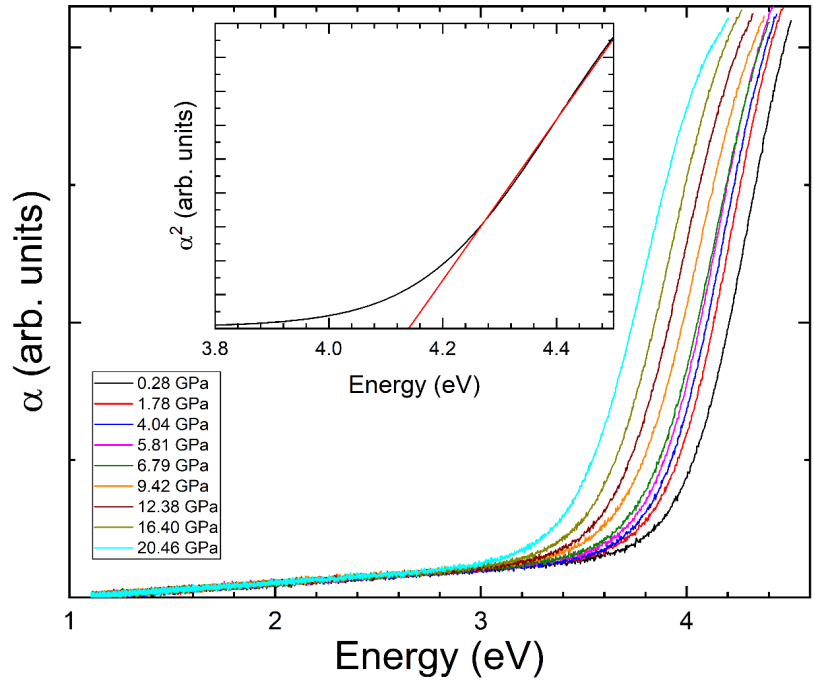

**Figure 9:** Absorbance ($\alpha$) of $Sr(IO_3)_2HIO_3$ at different pressures. The inset shows $\alpha^2$ versus energy for the spectrum measured at 0.28 GPa. The red line is the extrapolation used to determine the band-gap energy.

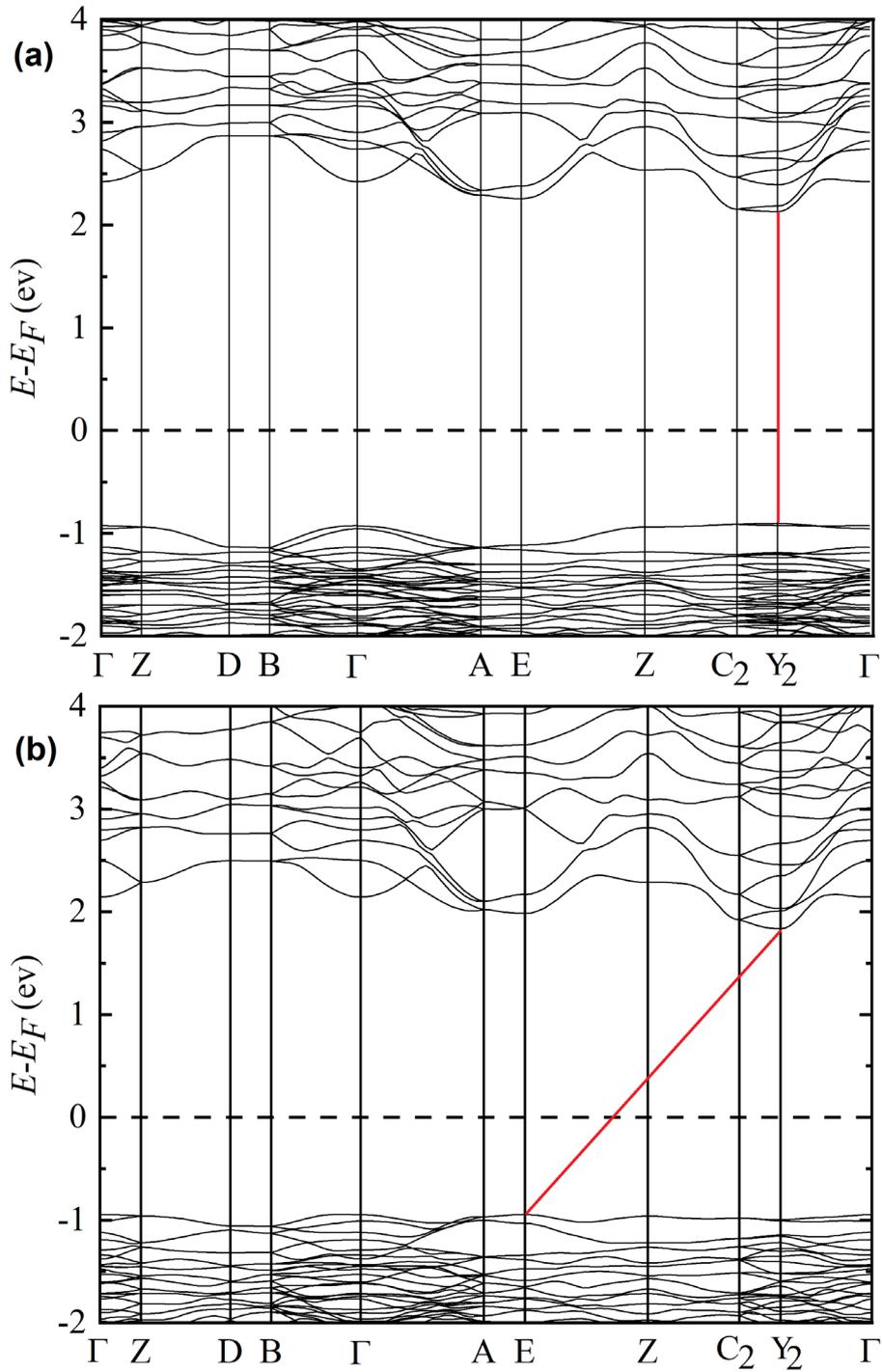

**Figure 10:** Calculated band structure of $Sr(IO_3)_2HIO_3$ at 0 GPa (a) and 7.32 GPa (b). Band gaps are identified by red lines.

Figures 10a and 10b show the band structure calculated at 0 GPa and 7.32 GPa. At 0 GPa the band gap is direct. The valence band maximum (VBM) and the conduction band minimum (CBM) are at the $Y_2$ point of the Brillouin zone. The calculated band-gap energy is 3.35 eV. This value underestimates the band gap determined from experiments. This is a typical issue of DFT calculations with the PBEsol functional [64], however, it does not prevent it from providing an accurate description of changes induced by pressure in the band-gap energy. Figure 10a shows that the valence band is not very dispersive with several maxima at similar energies. At 7.32 GPa the band gap is indirect with the CBM still at Y2, but the VBM at E.

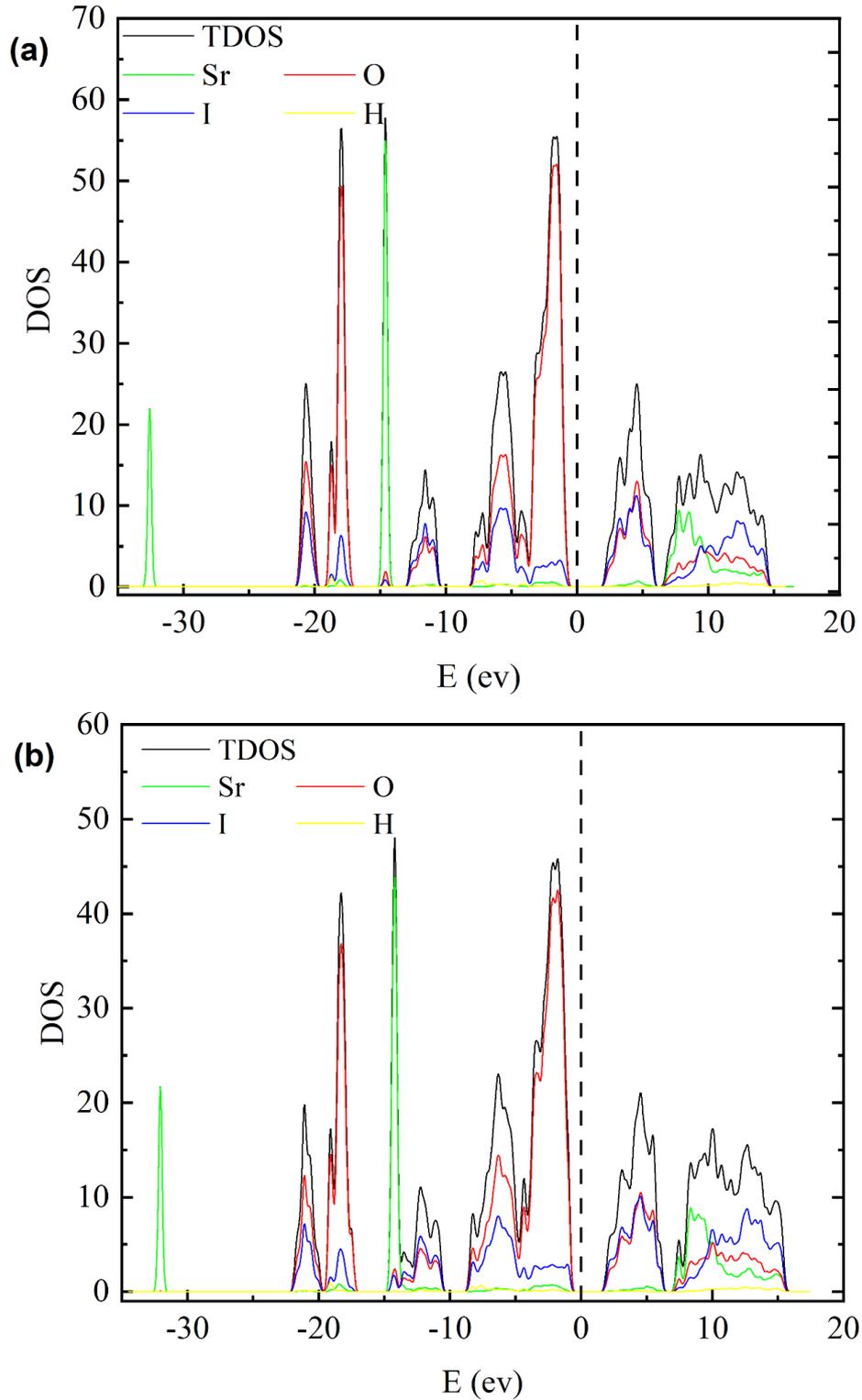

**Figure 11:** Total and partial electronic density of states of Sr(IO$_3$)$_2$HIO$_3$ calculated at 0 GPa (a) and 7.32 GPa (b).

To get a deeper understanding of the electronic properties of Sr(IO$_3$)$_2$HIO$_3$ we have calculated the electronic density of states (DOS). We show in Figures 11a and 11b the DOS at 0 and 7.32 GPa, respectively. We find that the VBM is dominated by O-2p orbitals with a small contribution of I-5p states. In addition, the CBM is dominated by equal contributions from O-2p and I-5p orbitals. At 7.32 GPa, despite the band-gap crossing, there are no noticeable changes in the composition of the VBM and CBM. The most evident change is only the narrowing of the band gap as pressure increases. This narrowing is a consequence of the enlargement under compression of I-O bonds within the IO$_3$ pyramids which cause a decrease in the splitting between bonding and antibonding states as observed in other iodates [14]. Regarding the orbital contributions, the DOS of Sr(IO$_3$)$_2$HIO$_3$ resembles those reported for other non-transition-metal iodates [1, 14]. The common characteristic of dominant contributions from O-2p and I-5p orbitals to VBM and CBM implies that the band-gap energy in metal iodates is strongly affected by the I-O interaction. According to Liang *et al.* [14] the band-gap energy, $E_{gap}$, of a metal iodate can be accurately estimated from the average I-O distance, $<d_{I-O}>$, by using the following relationship: $E_{gap} = -42.1(2.4) \times <d_{I-O}> + 80.4(4.4)$, where $E_{gap}$ is given in eV and $<d_{I-O}>$ is in Å. In this sense, we find the average of the covalent bonds of the three IO$_3$ pyramids of Sr(IO$_3$)$_2$HIO$_3$ to be 1.824 Å at RP. Using this value, one can obtain a value of 4.0(2) eV for the band-gap energy, which agrees with the present and previous experiments. Notice that according to present DFT calculations the average I-O distance is 1.848 Å. Using the model proposed by Liang *et al.* [14] the band gap is estimated to be 3.3(2) eV, which coincides with the value obtained from band-structure calculations.

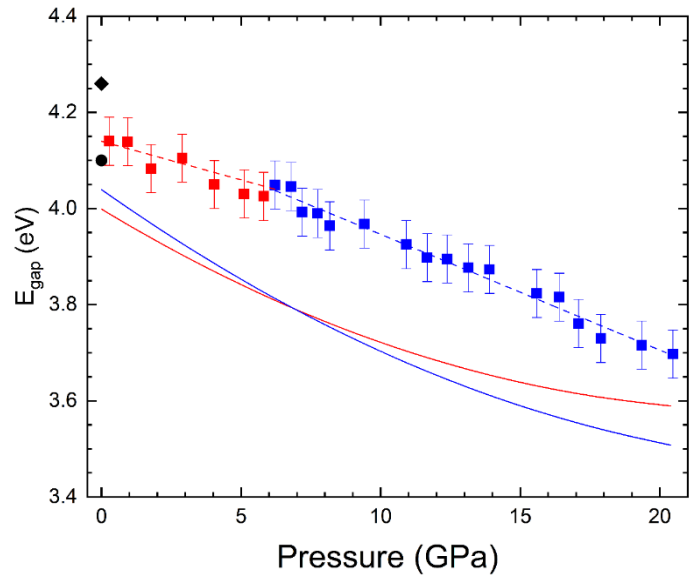

**Figure 12:** Pressure dependence of the band-gap energy of Sr(IO$_3$)$_2$HIO$_3$. Symbols are results from experiments. Solid lines are results from calculations and dashed lines are linear fits to the results of experiments. Red and blue are used for the direct and indirect gap, respectively. The black diamond (circle) are results from Ref. 19 (Ref. 26). The results of calculations have been shifted up by 0.7 eV to facilitate comparison with experiments.

To conclude this section, we discuss the pressure dependence of the band gap in Sr(IO$_3$)$_2$HIO$_3$. The results of our experiments and calculations are shown in Figure 12. According to our calculations there is a band crossing from a direct to an indirect gap near 7 GPa. This is because the dimer formation modifies the topology of the top of the valence band in a way that the VBM moves from the Y$_2$ point of the Brillouin zone to the E point. The band crossing implies a change in the slope of the pressure dependence of the band-gap energy, with the decrease of the band-gap energy being more pronounced for the indirect gap than for the direct gap. Notice that the decrease of the band-gap energy is in full agreement with the understanding achieved by Liang *et al.* [14] for the band-gap energy of iodates. In Figure 4e we show that from 0 to 20 GPa there is an increase of 0.02 Å in the average bond distance. Given the fact that the longer the average I-O bond length the smaller the band-gap energy, we expect a decrease of the band-gap energy with pressure. Using the relationship reported in Ref. 14 the decrease of the band-gap energy from 0 to 20 GPa is 0.5 eV, which is fully consistent with our results. In the experiments we found a qualitatively similar behavior than predicted by calculations. In particular, there is also a slope change in the pressure dependence of the band gap, in this case near 6 GPa. The slope gets steeper above this pressure, in agreement with calculations and supporting the occurrence of the band crossing.

**3.4 High-pressure effects on the refractive index**

Iodates form a class of compounds that possess unique bifrefringence properties which meade them ideal for non-linear optical applications [1]. Such properties coud be affected by changes in the crystal structure like those here reported. To bring light on this phenomenon we performed DFT calculations. Calculated spectra of the real and imaginary part of the dielectric function along the main crystallographic directions at different pressures are represented in Figure 13. From them we have calculated the refractive index for the three crystallographic axis. The results are reported in Figure 14. The static values of the refractive index (n) at 0 eV calculated along the main crystallographic directions at different pressures are presented in the Table 2. In the table it can be seen that our results show that the refractive index increases slightly with increasing the pressure. This is in agreement with what has already been seen in other materials in which an increase in the dielectric constant is caused by the hypercoordination related to the formation of electron-poor multicenter bonds that we have found in Sr(IO$_3$)$_2$HIO$_3$ [25]. Regarding birefringence in the visible range, the most notable change with pressure is that at 0 GPa the greatest difference in refractive indices occurs between n$_a$ and the n$_b$ and n$_c$ which are similar, but at 19 GPa it occurs between n$_c$ and the other two indexes. That is, the axe in which birefringence is manifested changes under compression, which is a consequence of pressure-induced bonding changes. Therefore, stresses can be an efficient tool for tuning birefringence for different applications.

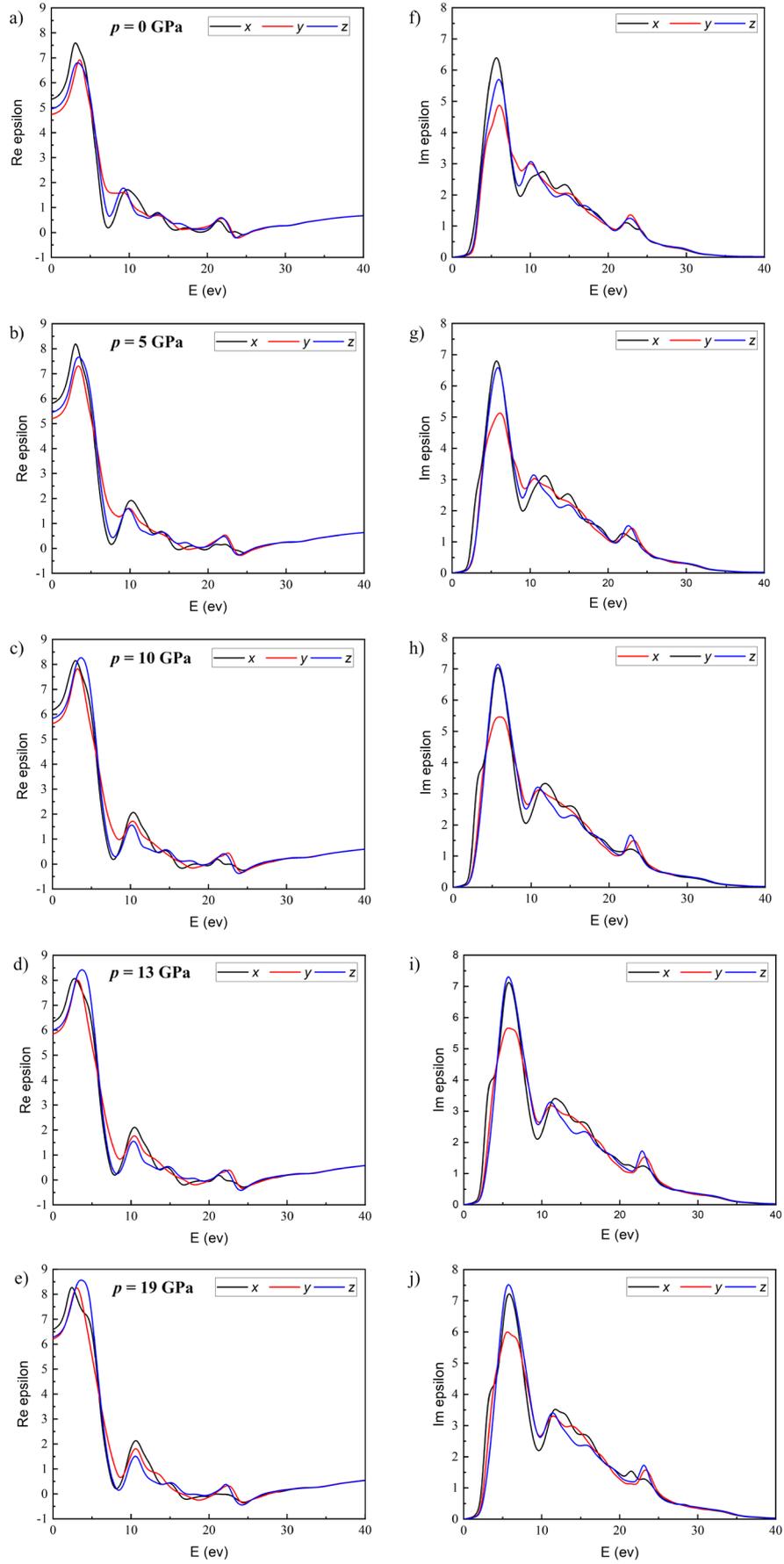

**Figure 13.** Components of the real, $\varepsilon_1$, (a-e) and imaginary, $\varepsilon_2$, (f-j) parts of the dielectric function as a function of energy for the Sr(IO$_3$)$_2$HIO$_3$ crystal at different pressures.

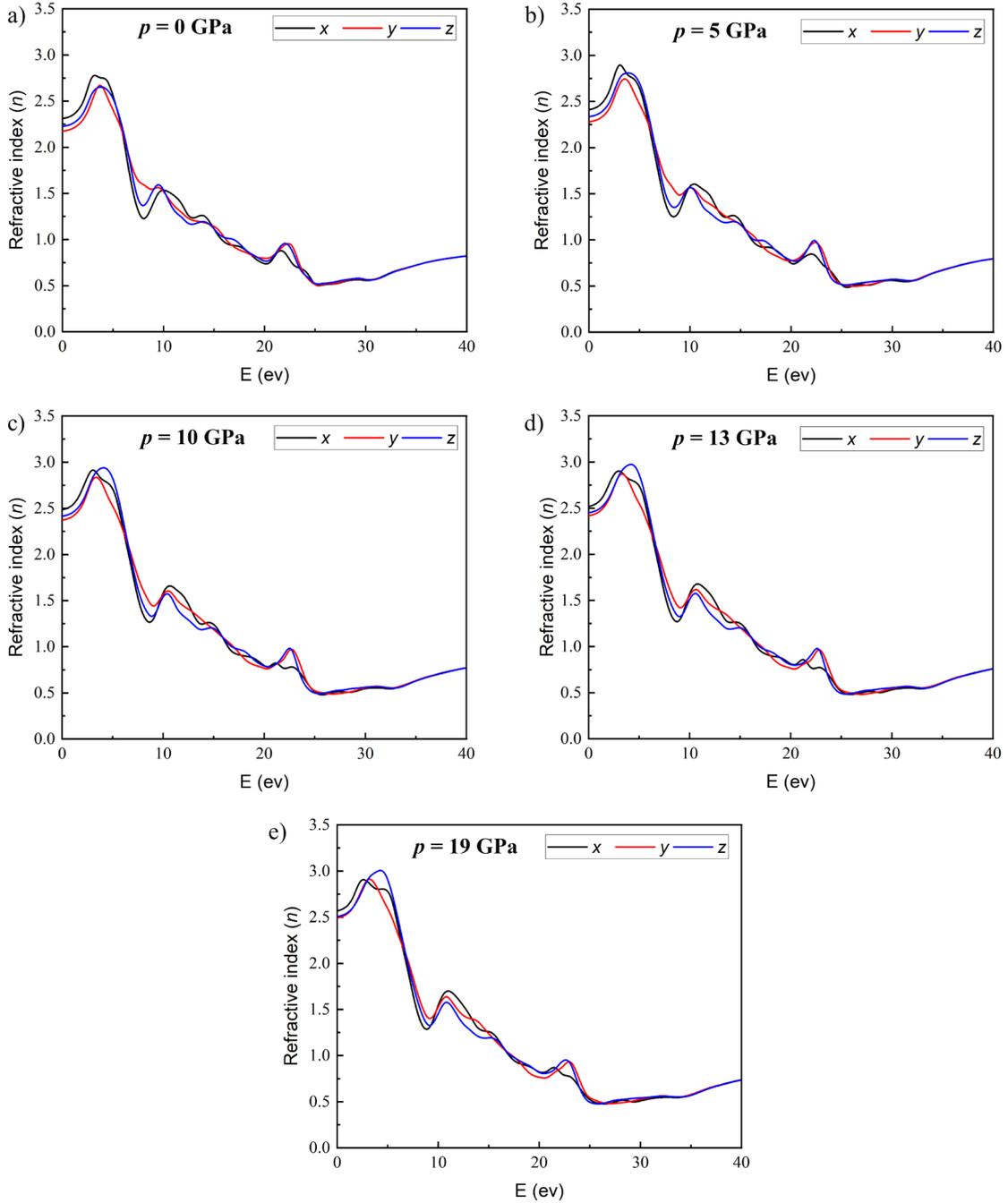

**Figure 14:** Refractive index along the main directions as a function of energy for the Sr(IO₃)₂HIO₃ crystal at different pressures.

## 4. Conclusions

In this work, we report synchrotron-based powder X-ray diffraction, Raman spectroscopy, and optical-absorption measurements performed up to 20 GPa in Sr(IO₃)₂HIO₃. The experiments have been combined with density-functional theory calculations and a computational study of the topology of the electron density. The compressibility of Sr(IO₃)₂HIO₃ has been studied and it has been found that this material is highly compressible with an anisotropic compressibility. The main finding of the study is the discovery of pressure-driven configurational changes that lead to iodine hypercoordination and subsequent dimerization. We provide evidence that pressure induces internal changes in the crystal structure favoring the linking between $HOI_3$ and $IO_3$ units. The changes happen in two steps. First $[HIO_3]\cdot[IO_3]$ complexes are formed around 2.5 GPa; then a $[HIO_3]\cdot[IO_3]\cdot[IO_3]\cdot[HIO_3]$ dimer is formed above 4.5 GPa. The changes observed in the crystal structure are a consequence of the transformation of several weak secondary (halogen) I···O bonds into a different type of bonds, the recently proposed electron-deficient multicenter bonds. This chemical transformation affects the crystal structure, the behavior of phonons, and the electronic band structure. In particular, an interesting phenomenon is the anomalous enlargement of some covalent I-O bonds that participate in the formation of the pressure-induced electron-deficient multicenter bonds. This phenomenon triggers the softening of the internal I-O vibrations of $IO_3$ units related to these enlarged bonds and the decrease of the band-gap energy, from 4.1 eV at 0 GPa to 3.7 eV at 20 GPa. In addition, the decrease of the band-gap energy causes a pressure-induced band-gap crossing. Finally, according to our calculations pressure affects the birefringence of the studied material changing the directions in which birefringence is modified, which is a consequence of pressure-induced bonding changes.

**Supporting Information**

Calculated Bader atomic charges, electrons transferred, and electrons shared. Calculated phonon frequencies including symmetry assignment and quadratic fits for the pressure dependence of phonons.

**CRediT authorship contribution statement**

D. Errandonea and F.J. Manjon: Conceptualization, Formal analysis, Supervision, Funding acquisition, Project administration, Writing - Original Draft, Writing - Review & Editing; H. Osman, R. Turnbull, D. Diaz-Anichtchenko, A. Liang, J. Sanchez-Martin, C. Popescu, D. Jiang, H. Song, and Y. Wang: Investigation, Writing - Review & Editing.

**Declaration of competing interest**

The authors declare that they have no known competing financial interests or personal relationships that could have appeared to influence the work reported in this paper.

**Acknowledgments**

The authors thank the financial support from the Spanish Ministerio de Ciencia e Innovación (https://doi.org/10.13039/501100011033) under Projects PID2019-106383GB-41/42, PID2022-138076NB-C41/42, and RED2022-134388-T. They also acknowledge the financial support of

Generalitat Valenciana through grants PROMETEO CIPROM/2021/075-GREENMAT, MFA/2022/007, and MFA/2022/025 (ARCANGEL). This study forms part of the Advanced Materials program and is supported by MCIN with funding from the European Union Next Generation EU (PRTR-C17.I1) and by the Generalitat Valenciana. J. S.-M. acknowledges the Spanish Ministry of Science, Innovation, and Universities for the PRE2020-092198 fellowship. R.T. acknowledges funding from the Generalitat Valenciana for Postdoctoral Fellowship no. CIAPOS/2021/20. C. P. acknowledges the financial support from the Spanish Ministerio de Ciencia e Innovacion through project PID2021-125927NB-C21. The authors thank ALBA for providing beamtime under experiment no. 2022085940 and the Universitat de Valencia for providing computational resources at the Tirant supercomputer from Red Española de Supercomputación. The authors thank constructive comments from anonymous reviewers.